\documentclass[preprint,12pt,3p]{elsarticle}

\usepackage{amssymb,amsmath,amsfonts}
\usepackage{graphicx}
\usepackage{bm}
\usepackage[T1]{fontenc}
\usepackage[utf8]{inputenc}
\usepackage{color}
\usepackage{caption}
\usepackage{subcaption}
\usepackage{url}
\usepackage{float}

\renewcommand{\vec}[1]{\bm{\mathrm{#1}}}

\def \F{\vec{F}}

\def \x{\vec{x}}
\def \us{u^\ast}
\def \ust{\tilde{u}^\ast}

\def\U {{{\bf u}}}
\def\F {{{\bf F}}}

\def\Y {{{\bf Y}}}

\def\Lap {\nabla^2}

\def\div {{\nabla \cdot}}

\journal{Fluids}

\graphicspath{{./figures/}}

\begin{document}

\begin{frontmatter}

\title{Three-dimensional low Reynolds number flows near biological filtering and protective layers}

%
%
\address[label1]{Department of Mathematics, University of Tennessee Knoxville, 1403 Circle Drive, Knoxville TN, 37996}
\ead{cstric12@utk.edu}
\address[label2]{Department of Mathematics, CB 3250, University of North Carolina, Chapel Hill, NC 27599}
\address[label3]{Department of Biology, CB , University of North Carolina, Chapel Hill, NC 27599}
\address[label4]{School of Mechanical and Aerospace Engineering, Oklahoma State University, 218 Engineering North, Stillwater, OK 74078}
\address[label5]{Department of Mathematics, Bucknell University, Lewisburg, PA 17837}
\address[label6]{Department of Mathematics and Statistics, The College of New Jersey, 2000 Pennington Rd., Ewing, NJ 08628}
\address[label7]{Army Research Office, 4300 S Miami Blvd, Durham, NC 27703}

%
%
\author[label1]{Christopher Strickland}
\ead{cstric12@utk.edu}
\ead[url]{https://www.christopherstrickland.info/}

\author[label2,label3]{Laura Miller\corref{cor1}}
\ead{lam9@unc.edu}
\ead[url]{http://miller.web.unc.edu}

\author[label4]{Arvind Santhanakrishnan}
\ead{askrish@okstate.edu}
\ead[url]{http://www.appliedfluidslab.org/}

\author[label5]{Christina Hamlet}
\ead{ch051@bucknell.edu}
\ead[url]{https://sites.google.com/site/chamlet5127/}

\author[label6]{Nicholas A. Battista}
\ead{nickabattista@gmail.com}
\ead[url]{http://nickabattista.wixsite.com/home}

\author[label2,label7]{Virginia Pasour}
\ead{virginia.b.pasour.civ@mail.mil}

\cortext[cor1]{I am corresponding author}

%
%
%
%

\begin{abstract}

Mesoscale filtering and protective layers are replete throughout the natural world. Within the body, arrays of extracellular proteins, microvilli, and cilia can act as both protective layers and mechanosensors. For example, blood flow profiles through the endothelial surface layer determine the amount of shear stress felt by the endothelial cells and may alter the rates at which molecules enter and exit the cells. Characterizing the flow profiles through such layers is therefore critical towards understanding the function of such arrays in cell signaling and molecular filtering. External filtering layers are also important to many animals and plants. Trichomes (the hairs or fine outgrowths on plants) can drastically alter both the average wind speed and profile near the leaf's surface, affecting the rates of nutrient and heat exchange. In this paper, dynamically scaled physical models are used to study the flow profiles outside of arrays of cylinders that represent such filtering and protective layers. In addition, numerical simulations using the Immersed Boundary Method are used to resolve the three-dimensional flows within the layers. The experimental and computational results are compared to analytical results obtained by  modeling the layer as a homogeneous porous medium with free flow above the layer. The experimental results show that the bulk flow is well described by simple analytical models. The numerical results show that the spatially averaged flow within the layer is well described by the Brinkman model. The numerical results also demonstrate, however, that the flow can be highly three-dimensional with fluid moving into and out of the layer. These effects are not described by the Brinkman model and may be significant for biologically relevant volume fractions. The results of this paper can be used to understand how variations in density and height of such structures can alter shear stresses and bulk flows. 

\end{abstract}

%
%
%
%


\end{frontmatter}



%
%
%
%

\section{Introduction}
\label{sec1}

Flows through porous layers are significant to numerous applications in biology, from the microscale problem of blood flow at the endothelial cell membrane to the macroscale problem of water flow through sea grass meadows and kelp forests \cite{Jackson83,Gaylord04}. Flows at the intermediate scales (e.g. the mesoscale) are also significant to many organisms because this is the scale at which many small organisms, such as tiny insects and zooplankton, experience their environment \cite{Koehl10,Koehl11,Jones:16}.

Bristled, hairy and filtering structures in flows where both viscosity and inertia are significant help many organisms feed, locomote, smell, signal, exchange nutrients, and perform a variety of sensory and other important and often complex functions.  These structures are found in many different types and sizes of organisms, in terrestrial and aquatic environments, and from the cellular level up to the organism level.  For example, cilia are hair-like protrusions that can act as ‘antennae’ on the surface of most eukaryotic cells  \cite{Ludeman}. Cilia also perform a variety of complex behaviors, including the establishment of left-right symmetry in the vertebrate embryo \cite{Babu}, acoustic streaming in the ear \cite{Lighthill}, and cell-cell communication in Drosophila \cite{Bornschlogl}.  Their presence and function in sponges, one of the earliest developing phyla, suggest that arrays of cilia such as these might represent the beginning of sensory and coordination systems in the animal kingdom \cite{Ludeman}.

Many examples also exist in the aquatic environment. For example, copepods beat their cephalic appendages to feed or swim, use antennules and swimming legs to escape from predators, or sink in the water column by reducing the use of their appendages \cite{Jiang2}.  Barnacles use their feeding structures, called cirral fans, to assist in their suspension feeding, sweeping them through the water when flow is slower and holding them out passively when currents are slower \cite{Geierman, Alexander}, and the lappets of developing scyphomedusa assist in propulsion as well as feeding \cite{Feitl, Wilson}.  On land, bristled, flexible wings help insects feed by lessening the force required to clap their wings together and fling the wings apart \cite{Miller:2009}.  Among mammals, lab rats possess different lengths of whiskers (vibrissae) that are attuned to different resonances, while the whiskers of pinnipeds (seals) allow them to follow prey by tracking hydrodynamic trails in the water \cite{Summarell}.

Another way in which small hairs aid organisms by altering the boundary layer over exchange surfaces. For example, trichomes are small hairs or other outgrowths that can be found on the epidermis of a plant during some or all stages of its life. One of the potential functions of the trichomes is to alter the boundary layer to increase the efficiency of gas exchange and transpiration. Turbulent boundary layers experience much higher rates of gas transport than do laminar boundary layers, but unless the windspeed is high, boundary layers can experience laminar flow even if the air around the plant is turbulent \cite{Vogel}.  Trichomes have been shown to increase the surface roughness of leaves, producing turbulent boundary layers at lower wind speeds and enhancing gas exchange \cite{Schreuder}.  Along with increasing the exchange of gases between plants and the atmosphere,  trichomes are also important for reducing absorbed solar radiation and leaf temperature, defending against predators, and reducing water film on leaves that could reduce gas exchange \cite{Schreuder}.



%
%
%
%

\subsection{Relevant dimensionless numbers and the leaky to solid transition}

To accurately describe flow through these filtering, protective, and hairy layers, several dimensionless parameters are needed. The first dimensionless number we will use is the bulk flow Reynolds number, $Re$. This Reynolds number describes the relative inertial to viscous forces in the bulk flow and is described by
\begin{equation}
    Re = \frac{\rho U L}{\mu}
\end{equation}
\noindent where $\rho$ is the viscosity of the fluid, $\mu$ is the dynamic viscosity of the fluid, $U$ is a characteristic velocity (here defined as the free stream velocity), and $L$ is a characteristic length scale. For the bulk flow $Re$, we use the height of the channel or computational domain as the characteristic length scale. To describe the relative importance of inertial and viscous effects at the level of the cylinders, we use a diameter based Reynolds number, $Re_d$, given by
\begin{equation}
    Re_d = \frac{\rho U D}{\mu}
\end{equation}
\noindent where $D$ is the diameter of the cylinder. Other important dimensionless numbers include the gap to diameter ratio, $G/D$. Here the diameter is set to the diameter of the cylinder, and the gap is set to the distance between the central axes of the nearest cylinders. The height to diameter ration, $H/D$, is set to the height of the cylinders to the diameter. Dimensionless quantities for several biological systems are given in Table \ref{bio-param}.

\begin{center}
\begin{table}
    \begin{tabular}{ | l | l | l | l | l | l | l | l |}
    \hline
    Structure  & Diameter & Height & Gap & G/D & H/D & $Re_d$ & Ref\\ \hline
    Glycocalyx & 10-12 nm & 150-400 nm & 20 nm & 2 & 12-40 & $\mathcal{O}(-3)$ & \cite{Wein:03}  \\ \hline
    Microvilli & 90 nm & 2.5 $\mu m$ & 165 nm & 1.83 & 28 & $\mathcal{O}(-3)$ & \cite{Guo}  \\ \hline
    Aesthetascs & 5.69-8.1 $\mu m$ & 347-648 $\mu m$ & - & 2-30 & 61-80 & $\mathcal{O}(-2)-\mathcal{O}(1) $ & \cite{Waldrop:13,Waldrop:14}  \\ \hline
    Bristled wings & 0.3-2.5 $\mu m$ & 25-200 $\mu m$ & 2-16 $\mu m$ & 5-10 & 10-150 & $\mathcal{O}(-2)$ &  \cite{Jones:16}  \\ \hline
    Trichomes   &28.1$\mu m$& 96.5$\mu m$ & 65.6$\mu m$  & 2.33  & 3.4 & $\mathcal{O}(1)$         & \cite{Szymanski:1999}  \\ \hline
    \hline
    \end{tabular}
    \caption{Measured morphological parameters of various cylinder-like structures in biology. The diameter based Reynolds Number, $Re_d$, was computed using the length scale as the diameter of the cylinders, kinematic viscosity of air or water. The characteristic velocity was chosen as the free stream velocity (approximately 1-2 $m/s$ for trichomes, the wind speed on an average day). Trichome measurements were taken from images using \cite{PlotDigitizer:2001,Hedrick:2008}.}
    \label{bio-param}
\end{table}
\end{center}

An important and complex function of bristles, hairs, and other bioarrays at the mesoscale is that they operate at the transition from leaky to functionally solid layers. When flow is viscous, due to the no-slip condition, a velocity gradient develops between the surface and freestream flow, with smaller or slower objects (lower $Re$) resulting in a thicker (boundary) layer of fluid.  The result is that rows of denser (or slower) cylinders operate as solid layers, while cylinders that are faster or farther apart function as leaky sieves.  

At the organism level, hairy appendages are typically leaky at $Re\sim 10^{-1}$ and higher, while they are paddle-like at $Re\sim 10^{-2}$ and lower \cite{Koehl3}. Koehl showed a transition between leaky- (sieve) and solid- (paddle) like behavior of an array of hairs based on factors such as the hair size, speed, and spacing \cite{Koehl, Koehl2, Koehl3}.  The relationship of the type of behavior to these factors is complex and nonlinear. For example, Loudon and Koehl found a complex relationship between wall nearness and Reynolds number \cite{Loudon}, and Jones \{textit{et al.} \cite{Jones:16} found that leakiness depends upon the angle of attack of the structure relative to the direction of flow. In addition, while some organisms exist in one regime all of their lives, others may exist in one regime at one life stage and transition to another at a later stage or quickly switch from one to the other as it becomes advantageous  \cite{Feitl, Blough}.


%
%

\subsection{Analytical porous models}

At the microscale ($Re<<1$) in the field of vascular transport, relatively simple, analytical models have been used to describe the profile of the blood flow above and through the endothelial surface layer (ESL). The endothelial surface layer consists of the glycocalyx and attached plasma proteins and has a significant role in proper cell function since it serves as a vasculoprotective layer, a mechanotransducer, and a molecular filter. Mathematical models of the flow through this layer have been to estimate volumetric flow rates and shear stresses within this layer and at the cell surface. One of the more popular models uses the Brinkman equation where the ESL is treated as a homogenized porous layer\cite{Brinkman:1949,  Wein:03, Dami:04, Smit:03, Vinc:08, Leid:08, Ferk:07}. For example, Weinbaum \textit{et al.}\cite{Wein:03} model the glycocalyx as a Brinkman layer and
calculate the value of the hydraulic conductivity using estimates of the volume fraction of core proteins and by assuming that the layer has a quasi-periodic structure. To obtain the flow profile for the entire vessel,
the flow through this layer is matched to Stokes flow above the layer. They find that the majority of shear stress is imposed on the tip of the core proteins and relatively little is imposed at the membrane. Brinkman models have also been used to describe the flow through layers of microvilli within the kidney that have a mechanosensory role \cite{Guo}.

At the larger scale ($Re>>1$), flow through vegetative canopies and beds, including sea grasses, reefs, and macrophytes, have been modeled using 1D models with analytical solutions such as Darcy's law \cite{HDarcy56}, its generalization to the Brinkman law \cite{Brinkman:1949}, and the Darcy-Forchheimer law \cite{Bejan} describing non-Darcy flow at high velocities. In some applications, modified Brinkman equations have been applied used to model flow through vegetation using a Cantor-Taylor brush configuration \cite{Shavit02,Shavit04}. In this case, three regions of the flow are considered: 1) a dense layer, 2) a sparse layer, and 3) a layer free of obstructions. The flow fields obtained from these modeling efforts have then been used to understand the flushing of plankton and pollen from sea grass meadows and kelp forests \cite{Jackson83,Gaylord04} and to describe the distributions of zooplankton that use local flow velocity and sheer to initiate reorientation or directed movements \cite{Grunbaum,Reidenbach:09,Koehl:10}. For a more complete overview of applications, please see the review by Nepf \textit{et al.} \cite{Nepf}.


By design, the Brinkman and similar porous models only describes the average flow through a porous layer and do not reveal smaller scale flow patterns where the fluid bends around obstacles such as various protrusions, transmembrane proteins, bristles, and hairs. Models that describe shear flow or flow within a channel with a porous layer typically assume that there is no flow in the third dimension (e.g. into and out of the layer). For applications where such small scale flow structures are important, it may be necessary to consider more than a model of the averaged flow. Example applications could include anything where a particle is required to diffuse or swim through such a layer. Furthermore, the error introduced by the homogenization of the porous layer will be proportional to the amount of flow in the third dimension. Careful consideration of this amount of flow in the third dimension as a function of obstacle spacing, $Re$, and height of the layer has not been carefully described, although studies of specific applications do suggest its importance \cite{Cheer:12, Jones:16, Waldrop:16}.

%
%

\subsection{Fully resolved flow (e.g., not averaged) past 3D structures}

Our goal in this paper is to describe the details of flow within idealized biological filtering, protective, and other types of porous layers, and solve an associated fully coupled fluid-structure interaction problem for simplified structures. While numerous studies have considered such flow at small ($Re<<1$) or large ($Re>>1$) scales, few studies have focused on the mesoscale ($10^{-2}<Re<10^1$). Different length and time scales and far-from-equilibrium spatially and temporally varying forces have motivated new mesoscale modeling approaches for studying the dynamics of hairy structures in fluids \cite{Winkler}. Another method that has proven useful is that of Cheer and Koehl; Stokes low-Re approximation is used to model velocities close to the hairs (cylinders) and Oseen’s low-Re approximation of the Navier Stokes equation is used far from the hairs. A matched asymptotic expansion technique is then applied in the areas between the hairs and outside of the array \cite{Cheer}.  


In this paper, we use several approaches to describe the flow around filtering layers at the mesoscale:\\
1) Flow around physical models of filtering layers is measured using particle image velocimetry. \\
2) A 1D Brinkman model of flow through porous layers is compared to three-dimensional dynamically scaled physical models. The goal is to confirm that the 1D Brinkman model captures bulk flow outside of the porous layers.\\
3) Three-dimensional flow through idealized filtering layers is numerically simulated using the immersed boundary method. \\
4) A 1D Brinkman model of flow within the layer is compared to the numerical simulations. The goal is to confirm that the Brinkman model adequately captures average flow but does not capture movement in the third dimension (which would enhance exchange into and out of the layer).\\

Note that the models and numerical simulations are also used to examine how changes in volume fraction, height, and arrangement of the filtering layer alters velocity profiles within and around the layer. Two different boundary conditions are considered for experimental and numerical feasibility. In the case of physical models, flow is driven through a channel with a porous layer on both the top and bottom. This is representative of flow through blood vessels and other internal channels. In the numerical simulations, shear flow is considered with a porous layer on the bottom of the domain. This is representative of flow over a leaf or another external surface. It is also a reasonable approximation of flow near the luminal surface of a vessel.

%
%
%
%

\section{Methods}
\label{sec2}

%
%
%
%

\subsection{Immersed Boundary Method}
\label{subsec1}

\begin{table}
\begin{center}
\begin{tabular}{| c | c |}
    \hline
    Parameter & Value \\ \hline
    $L$             &  1.0 m              \\ \hline
    $dx$            &  $L/512$ m          \\ \hline
    $ds$            &  $L/1024$ m        \\ \hline
    $dt$            &  $10^{-4}$ s        \\ \hline
    $\rho$        & 1000 $kg/m^3$       \\ \hline
    $\mu$         & \textit{varied}  \\ \hline
    $V$           & 0.1 $m/s$         \\ \hline
    $k_{targ}$    & $3.186\times10^{2}\ kg/s^2$ \\ \hline
    tower spacing & $L/8-L$  \\ \hline
    end time      & 10-200 s        \\ \hline
    $Re$      & 0.1-10       \\ \hline
    $Re_d$      & $3.124 \times 10^{-3}-10^{-1}$      \\ \hline
    $G/D$      & 4-32      \\ \hline
    $H/D$      & 5-20      \\ \hline
    \hline
    \end{tabular}
    \caption{Numerical parameters used in the three-dimensional simulations.}
    \label{num_param}
    \end{center}
\end{table}

To perform the fluid-structure interaction, we implement the immersed boundary method. The immersed boundary method has been successfully applied to numerous different applications from microscale and cellular interactions \cite{Atzberger:2007,Leiderman:2008,Strychalski:2013} to cardiovascular dynamics \cite{Peskin:1996,Griffith:2012}, to organismal scale biomechanics, including aquatic locomotion \cite{Fauci:1988,Hoover:2015}, insect flight \cite{Miller:2009,SJones:2015}, muscle-fluid-structure interactions \cite{Battista:2015,Hamlet:2015}, and plant biomechanics \cite{Zhu:2011,Miller:2012}, to parachuting \cite{Kim:2006}. 

The following outlines the two-dimensional formulation of the immersed boundary method, from which the three dimensional extension is straightforward. For a full review of the method, please see Peskin \cite{Peskin:2002}. The governing equations of fluid motion are given by the Navier-Stokes equations:

\begin{eqnarray} 
   \rho\left[\frac{\partial\U}{\partial t}({\bf x},t) +\U({\bf x},t)\cdot\nabla \U({\bf x},t)\right] &=&  \nabla p({\bf x},t) + \mu \Delta \U({\bf x},t) + \F({\bf x},t) \label{eq:NS1} \\
  \div \U({\bf x},t) &=& 0 \label{eq:NSDiv1}
\end{eqnarray}
where $\U({\bf x},t) $ is the fluid velocity, $p({\bf x},t) $ is the pressure, $\F({\bf x},t) $ is the force per unit area applied to the fluid by the immersed boundary, $\rho$ and $\mu$ are the fluid's density and dynamic viscosity, respectively. The independent variables are the time $t$ and the position ${\bf x}$. Note that the variables $\U, p$, and $\F$ are all written in an Eulerian framework on the fixed Cartesian mesh, $\textbf{x}$. 

The interaction equations between the fluid and the boundary are given by:
\begin{eqnarray}
   {\bf F}({\bf x},t) &=& \int {\bf f}(q,t)  \delta\left({\bf x} - {\bf X}(q,t)\right) dq \label{eq:force1} \\
   {\bf X}_{t}(q,t) = {\bf U}({\bf X}(q,t)) &=& \int \U({\bf x},t)  \delta\left({\bf x} - {\bf X}(q,t)\right) d{\bf x} \label{eq:force2}
\end{eqnarray}
where ${\bf f}(q,t)$ is the force per unit length applied by the boundary to the fluid as a function of Lagrangian position, $q$, and time, $\delta({\bf x})$ is a two-dimensional delta function, ${\bf X}(q,t)$ gives the Cartesian coordinates at time $t$ of the material point labeled by the Lagrangian parameter, $q$.  Equation (\ref{eq:force1}) applies a singular force from the immersed boundary to the fluid through the external forcing term in Eq.(\ref{eq:NS1}), and Eq.(\ref{eq:force2}) evaluates the local fluid velocity at the boundary. This enforces the no-slip condition and the boundary is then moved at the local fluid velocity. Each interaction equation involves an integral transformation, with a two-dimensional Dirac delta function kernel, $\delta$,  to convert Lagrangian variables to Eulerian variables and vice versa.

The forcing term, $\textbf{f}(q,t)$, in the integrand of Eq.(\ref{eq:force1}) is specific to the application. In a simple case where boundary points are tethered to target points, to hold the boundary nearly fixed, the equation describing the force applied to the fluid by the boundary is
\begin{equation}
{\bf f}(q,t) = k_{targ} \left(\Y(q,t) - {\bf X}(q,t)\right)
\label{eq:force3}
\end{equation}
where $k_{targ}$ is a stiffness coefficient and $\Y(q,t)$ is the prescribed position of the target boundary. Details on other forcing terms can be found in \cite{BattistaIB2d:2016,BattistaIB2d:2017}.

%
%
%
%

\subsection{Description of the numerical setup and example output}

We use the immersed boundary method \cite{Peskin:2002} to simulate the fully-coupled fluid structure interaction (FSI) problem of a single cylinder in a periodic flow. Fluid equations are nondimensionalized and discretized on a Cartesian grid, and boundary equations are discretized on a moving Lagrangian grid. The full Navier-Stokes equations are solved for the fluid domain. We use IBAMR, a software library that provides an adaptive and parallelized framework for the direct numerical simulation of the FSI problem using the immersed boundary method \cite{BGriffithIBAMR}.

IBAMR is a C++ framework that provides discretization and solver infrastructure for partial differential equations on block-structured locally refined Eulerian grids \cite{MJBerger84,MJBerger89} and on Lagrangian meshes. Adaptive mesh refinement (AMR) is capable of higher accuracy between the immersed boundary and the fluid by increasing grid resolution in areas of the domain where vorticity exceeds a certain threshold or that contains an immersed structure. AMR improves computational efficiency by decreasing grid resolution in areas of the domain where low resolution is sufficient.

The Eulerian grid was locally refined near both the immersed boundaries and regions of vorticity where $|\omega| > 0.50$. This Cartesian grid was organized as a hierarchy of four nested grid levels; the finest grid was assigned a resolution of $dx = L/512$. A 1:4 spatial step size ratio was used between each successive grid level. The Lagrangian spatial step resolution was chosen to be twice the resolution of the finest Eulerian grid, e.g., $ds =\frac{L}{1024}$.

All numerical parameters are provided in Table \ref{num_param}. The height of the domain was set to 1.0, and the lengths in the x- and y-directions were set to $1/8$. The domain is periodic in the $x$- and $y$ directions. The boundary conditions are set in the $z$-direction to model shear flow. The $x$-component of the velocity was set to $0.1\ m/s$ at the top of the domain and to $0\ m/s$ at the bottom. The diameter of the cylinder was set to $0.03125\ m$, and the height was varied from $0.078m$ to $0.625m$. Reynolds number (Re) was varied by changing the dynamic viscosity.

\begin{figure}[H]
    \centering
    \includegraphics[width=0.4\textwidth]{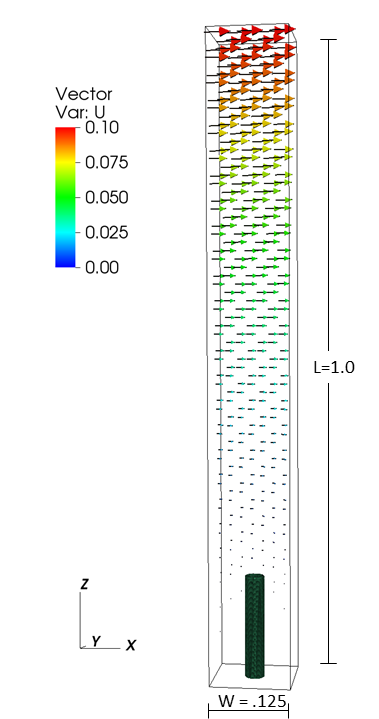}
    \caption{Numerical set up for the periodic cylinder configuration.}
    \label{fig:num_setup}
\end{figure}

%
%
%
%

\subsection{1D Analytical Model using Brinkman Equations}
\label{subsec2}

\begin{figure}[H]
    \centering
    \includegraphics[width=0.80\textwidth]{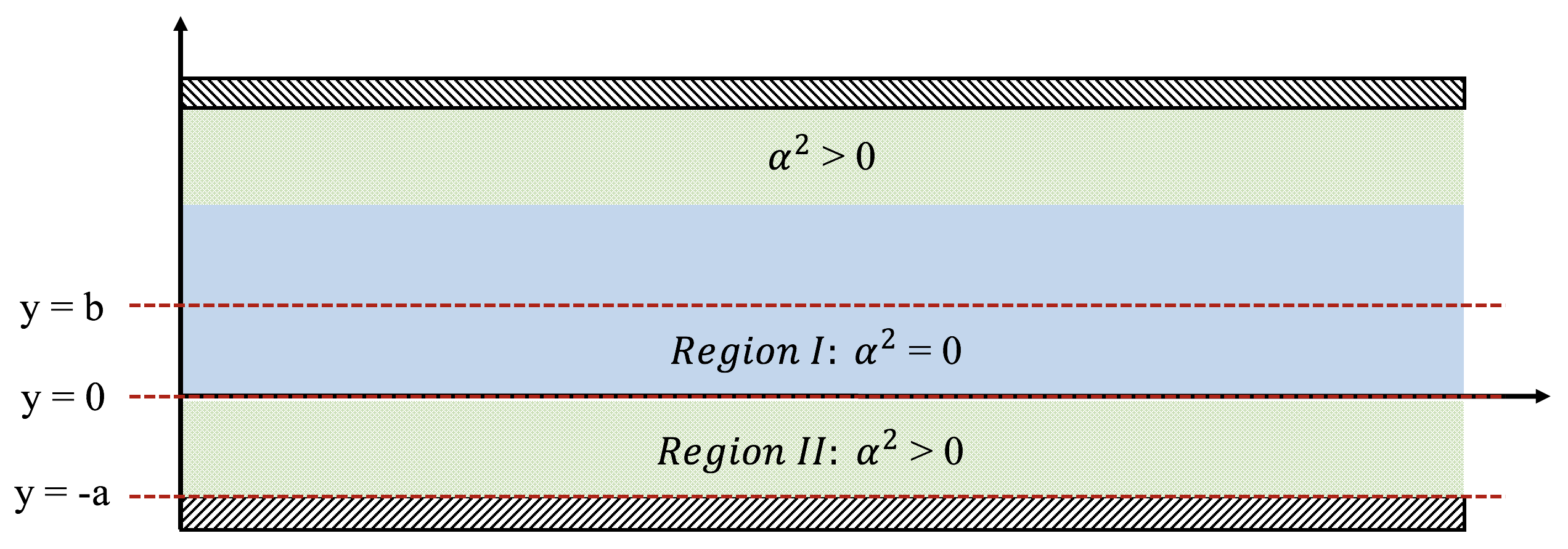}
    \caption{Depicting the physical setup for the Brinkman equations. There are two regions of interest, namely where $\alpha^2=0$ and $\alpha^2>0$. }
    \label{fig:brinkman_setup}
\end{figure}

Brinkman's equation, a generalization of Darcy's law, states that the force acting on the fluid due to a porous layer is proportional to the local velocity in the layer\cite{Brinkman:1949}. Brinkman's equation is written as
\begin{equation}
  \rho(\U_{t}({\bf x},t) +\U({\bf x},t)\cdot\nabla \U({\bf x},t)) =  -\nabla p({\bf x},t) + \mu \Lap \U({\bf x},t) - \alpha \mu \U({\bf x}, t)
  \end{equation}
where is $\alpha$ is the inverse of the hydraulic permeability. 


First, we rewrite the two-dimensional Brinkman equation as
\begin{align}
\frac{\partial u}{\partial t}+u\frac{\partial u}{\partial x}+v\frac{\partial u}{\partial y} &= -\frac{1}{\rho}\frac{\partial p}{\partial x} + \eta\left(\frac{\partial^2 u}{\partial x^2} + \frac{\partial^2 u}{\partial y^2}\right) - \frac{\alpha^2}{\rho}\mu u\label{eqn:Brink1}\\
\frac{\partial v}{\partial t}+u\frac{\partial v}{\partial x}+v\frac{\partial v}{\partial y} &= -\frac{1}{\rho}\frac{\partial p}{\partial y} + \eta\left(\frac{\partial^2 v}{\partial x^2} + \frac{\partial^2 v}{\partial y^2}\right) - \frac{\alpha^2}{\rho}\mu v\label{eqn:Brink2}
\end{align}
where $\eta = \mu/\rho$. We consider a channel consisting of a central main region with unobstructed parallel shear flow (Region I with $\alpha^2=0$) that is bounded on the top and the bottom by identical porous layers (Region II with $\alpha^2 > 0$). The flow field is assumed to be symmetric about the centerline $y=b$, and so our analysis will be restricted to the bottom half of the system. The flow is considered to be steady ($\partial u/\partial t = 0$), fully developed $(\partial u/\partial x=0$), and zero in the cross-stream direction ($v=0$).

%
%

\paragraph{Region I: } Using these assumptions, the $x$-momentum equation (\ref{eqn:Brink1}) in the free shear flow region (Region I, where $\alpha^2=0$) simplifies to
\begin{equation}
\frac{d^2u}{dy^2}=-\frac{1}{\mu}\frac{\partial p}{\partial x}.\label{eqn:Reg1Brink}
\end{equation}
The $y$-momentum equation (\ref{eqn:Brink2}) simplifies to
\begin{equation*}
\frac{\partial p}{\partial y} = 0 \Rightarrow p=p(x).
\end{equation*}
In other words, the fluid pressure is independent of the $y$-direction, and only changes in the $x$-direction.

We are interested in determining a closed-form expression for the velocity of the flow field within this region. To obtain this, equation (\ref{eqn:Reg1Brink}) is integrated once in $y$ to obtain the velocity gradient
\begin{equation}
\frac{du}{dy} = \frac{y}{\mu}\frac{dp}{dx} + A\label{eqn:Reg1dersol}
\end{equation}
and then integrated once again to arrive at an equation for the streamwise velocity
\begin{equation}
u(y) = \frac{y^2}{2\mu}\frac{dp}{dx} + Ay + B\label{eqn:Reg1gensol}
\end{equation}
where the evaluation of constants $A$ and $B$ requires two boundary conditions.

%
%

\paragraph{Region II: } To distinguish between the velocity of the flow in Region I, $\us$ will be used to denote the streamwise $x$-directional velocity in the porous layer. Starting with the equation set (\ref{eqn:Brink1})-(\ref{eqn:Brink2}) and using the same assumptions as in Region I with $\alpha^2>0$ gives
\begin{equation}
\frac{d^2\us}{dy^2} = \frac{1}{\mu}\frac{dp}{dx}+\alpha^2\us\label{eqn:Reg2Brink}.
\end{equation}
This second order ordinary differential equation can be converted into a set of two first order differential equations by defining a new variable $w$,
\begin{equation}
\frac{d\us}{dy}=w\label{eqn:wdef}
\end{equation}
such that equation (\ref{eqn:Reg2Brink}) can be reduced to
\begin{equation}
\frac{dw}{dy} = \frac{1}{\mu}\frac{dp}{dx}+\alpha^2\us\label{eqn:dw}.
\end{equation}
The equation set (\ref{eqn:wdef}) and (\ref{eqn:dw}) represents a two-dimensional system of non-homogenous, first order differential equations which can be further simplified into a homogenous system by defining a transformation for $\us$,
\[\ust = \left(\us + \frac{1}{\alpha^2\mu}\frac{dp}{dx}\right)\]
so that
\begin{align*}
\frac{d\ust}{dy} &= \frac{d\us}{dy} = w\\
\frac{dw}{dy} &= \alpha^2\ust.
\end{align*}
This system can be written in matrix form as
\[\frac{d}{dt}\left(
                \begin{array}{c}
                  \ust \\
                  w \\
                \end{array}
              \right) = \left(
                          \begin{array}{cc}
                            0 & 1 \\
                            \alpha^2 & 0 \\
                          \end{array}
                        \right)\left(
                                 \begin{array}{c}
                                   \ust \\
                                   w \\
                                 \end{array}
                               \right)
\]
and solved using conventional linear algebra techniques to give the velocity profile in Region II,
\begin{align}
\ust(y) &= Ce^{\alpha y}+De^{-\alpha y}\nonumber\\
\us(y) &= Ce^{\alpha y}+De^{-\alpha y} - \frac{1}{\alpha^2\mu}\frac{dp}{dx}\label{eqn:Reg2gensol}\\
w(y) &= \alpha Ce^{\alpha y} - \alpha De^{-\alpha y},\nonumber
\end{align}
where the constants $C$ and $D$ will be evaluated using the boundary conditions.

At the centerline of the flow domain $y=b$, velocity is unchanging
\[\left.\frac{du(b)}{dy}\right|_{Region\ I} = 0.\]
Applying this condition to equation (\ref{eqn:Reg1dersol}), we have
\begin{equation}
A = -\frac{b}{\mu}\frac{dp}{dx}.\label{eqn:ub_orig}
\end{equation}
At $y=-a$ in Region II, the layer of fluid that is in contact with the channel wall remains at rest (``no slip''). Applying $\us(-a)=0$ to equation (\ref{eqn:Reg1dersol}) yields
\begin{equation}
Ce^{-\alpha a}+De^{\alpha a} = \frac{1}{\alpha^2\mu}\frac{dp}{dx}\label{eqn:noslip}.
\end{equation}
Between the regions at $y=0$, matching of the velocities $u(0) = \us(0)$ results in
\begin{equation}
B = C+D-\frac{1}{\alpha^2\mu}\frac{dp}{dx}\label{eqn:B}
\end{equation}
and matching the velocity gradients gives us
\begin{equation}
A = \alpha C - \alpha D\label{eqn:gradmatch}.
\end{equation}
Solving equations (\ref{eqn:ub_orig}), (\ref{eqn:noslip}), (\ref{eqn:B}), and (\ref{eqn:gradmatch}), we have the constants
\begin{gather*}
    B = \frac{1}{\alpha^2\mu}\frac{dp}{dx}\left[\frac{2+e^{-\alpha a}(\alpha b-1) - e^{\alpha a}(\alpha b+1)}{e^{\alpha a}+e^{-\alpha a}}\right]\\
    C = \frac{1}{\alpha^2\mu}\frac{dp}{dx}\left[\frac{1-\alpha b e^{\alpha a}}{e^{\alpha a}+e^{-\alpha a}}\right]\\
    D = \frac{1}{\alpha^2\mu}\frac{dp}{dx}\left[\frac{1+\alpha b e^{-\alpha a}}{e^{\alpha a}+e^{-\alpha a}}\right]
\end{gather*}
with $A$ given in equation (\ref{eqn:ub_orig}).

Substitution of these constants into equations (\ref{eqn:Reg1gensol}) and (\ref{eqn:Reg1dersol}) gives us the velocity distributions as
\paragraph{Region I:}
\begin{equation*}
u(y) = \frac{1}{\alpha^2\mu}\frac{dp}{dx}\left(\frac{1}{2}\alpha^2y^2-b\alpha^2y + \frac{2+e^{-\alpha a}(\alpha b-1) - e^{\alpha a}(\alpha b+1)}{e^{\alpha a}+e^{-\alpha a}}\right)
\end{equation*}
\paragraph{Region II:}
\begin{equation*}
\us(y) =\frac{1}{\alpha^2\mu}\frac{dp}{dx}\left(\frac{\alpha be^{-\alpha(y+a)}-\alpha be^{\alpha(y+a)} + e^{-\alpha y} + e^{\alpha y}}{e^{\alpha a}+e^{-\alpha a}} - 1\right)
\end{equation*}
The average flow velocities can be determined using the integral definitions
\paragraph{Region I}
\begin{equation*}
\bar{u} = \frac{1}{b}\int_0^b u(y)\ dy = \frac{1}{\alpha^2\mu}\frac{dp}{dx}\left[\frac{2+e^{-\alpha a}(\alpha b-1) - e^{\alpha a}(\alpha b+1)}{e^{\alpha a}+e^{-\alpha a}} - \frac{1}{3}\alpha^2b^2\right]
\end{equation*}
\paragraph{Region II}
\begin{equation*}
\bar{u}^\ast = \frac{1}{a}\int_{-a}^0 \us(y)\ dy = \frac{2}{\alpha^3\mu a}\frac{dp}{dx}\left[\frac{\alpha b + e^{\alpha a}}{e^{\alpha a}+e^{-\alpha a}} - \frac{1}{2}(\alpha a + \alpha b + 1)\right] 
\end{equation*}
The viscous fluid flow exerts a tangential shear stress which can be determined as the gradient of the streamwise velocity, given as
\paragraph{Region I}
\begin{equation*}
\tau(y)=\frac{du}{dy} = \frac{1}{\mu}\frac{dp}{dx}(y-b)
\end{equation*}
\paragraph{Region II}
\begin{equation*}
\tau^\ast(y)=\frac{d\us}{dy} = \frac{1}{\alpha^2\mu}\frac{dp}{dx}\left[\frac{\alpha(e^{\alpha y} - e^{-\alpha y}) - \alpha^2b\left(e^{\alpha(y+a)}+e^{-\alpha(y-a)}\right)}{e^{\alpha a}+e^{-\alpha a}}\right]
\end{equation*}

We also consider the case of simple shear flow where velocity is a positive constant at $y=b$ ($U(b) > 0$) and a no-slip condition applies at the bottom ($u(-a)=0$), see Figure \ref{fig:brinkman_setup}. In this case, equation (\ref{eqn:ub_orig}) is replaced by
\begin{equation}
\frac{b^2}{2\mu}\frac{dp}{dx}+Ab+B=U>0.\label{eqn:ub}
\end{equation}
Solving equations (\ref{eqn:ub}), (\ref{eqn:noslip}), (\ref{eqn:B}), and (\ref{eqn:gradmatch}) for the constants $C$ and $D$ results in
\begin{align*}
C &= \frac{\frac{dp}{dx}(-\frac{1}{2}\alpha^2b^2+\alpha be^{-\alpha a}-e^{-\alpha a}+1)+U\alpha^2\mu}{ \alpha^2\mu \left(\alpha b e^{-2\alpha a}+\alpha b - e^{-2\alpha a}+1\right)}\\
D &= \frac{\frac{dp}{dx}(\frac{1}{2}\alpha^2b^2+\alpha be^{\alpha a}+e^{\alpha a}-1)-U\alpha^2\mu}{ \alpha^2\mu \left(\alpha b e^{2\alpha a}+\alpha b + e^{2\alpha a}-1\right)}.
\end{align*}
The other constants can be found in terms of these via equations (\ref{eqn:B}) and (\ref{eqn:gradmatch}) and then applied to equations (\ref{eqn:Reg1gensol}) and (\ref{eqn:Reg1dersol}) for the solution. In performing these calculations on a computer, it is useful to note that multiplying both the numerator and denominator of $D$ by $e^{-2\alpha a}$ results in an expression that shares the same denominator as $C$ while also avoiding numerical overflow in domains with large porous regions or values of $\alpha$.

%
%
%
%

\subsection{Physical Models}
\label{physical_model}

\begin{figure}[H]
\centering{\includegraphics[width=4in]{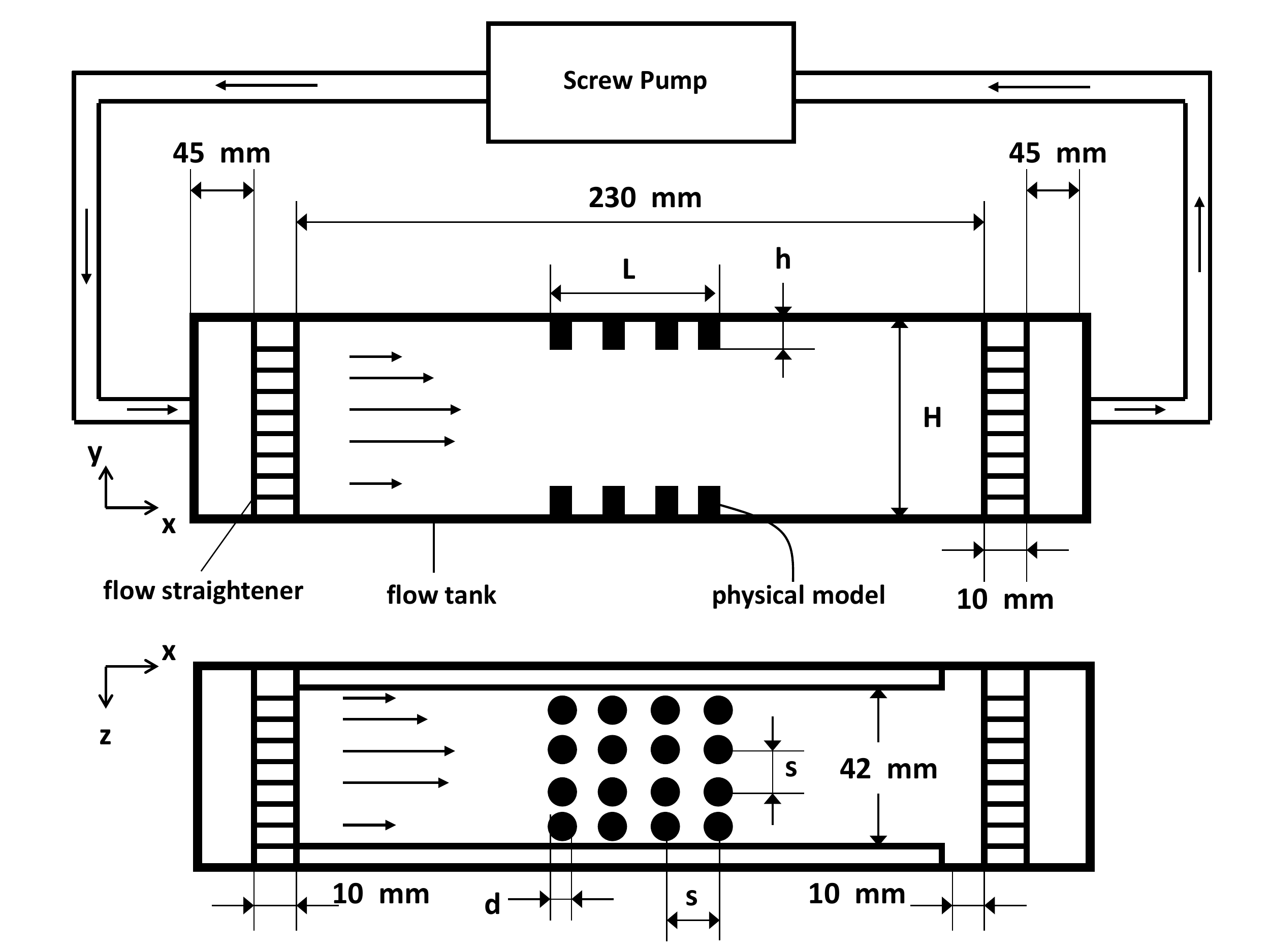}}
\caption{Schematic of the experimental setup showing the side view (top half) and the top view (bottom half) of the flow chamber with the physical model inserted. Flow direction in the tank is from right to left.}
\label{setup}
\end{figure}

The dynamically scaled physical models consisting of rigid cylinders are constructed out of pins that are 1 mm in diameter. Figure \ref{setup} shows the side and top views of the physical models. The parameters vary for each of the models, and the entire list of design parameters examined herein is presented in Table \ref{exp_param}. This setup is inspired by flow through the endothelial surface layer within capillaries. The $Re_d$ of the flow in capillaries is $\cal{O}$(10$^{-3}$), using the vessel diameter as the characteristic length and the maximum velocity as the characteristic velocity. The $Re_d$ of the flow within the endothelial surface layer is $\cal{O}$(10$^{-6}$), using the core protein diameter as the characteristic length scale and the maximum velocity in a capillary as the characteristic velocity. 

\begin{table}
\begin{center}
\begin{tabular}{| c | c |}
    \hline
    Parameter & Value \\ \hline
    $H$          &  $0.05\ m$ \\ \hline
    $d_{pin}$    &  $0.001\ m$              \\ \hline
    $V_{exp}$    &  $0.002\ m/s$           \\ \hline
    $\mu_{exp}$  &  $1.229 kg/(ms)$         \\ \hline
    $\rho_{exp}$ &  $1340\ kg/m^3$        \\ \hline
    layer length &  $\{0.01,0.043\} m$       \\ \hline
    layer height &  $\{0.008,0.022,0.028\} m$ \\ \hline
    pin spacing  &  $\{0.0025, 0.005, 0.01\}\ m$               \\ \hline
    $Re_d$       & $\sim 0.001$         \\ \hline
    $Re$         & $\sim 0.01$                \\ \hline
    \hline
    \end{tabular}
    \caption{Parameters used in the physical model experiments.}
    \label{exp_param}
    \end{center}
\end{table}

To achieve this low value of $Re$ in these experiments with macroscopically scaled models, both the velocity and the viscosity of the fluid are adjusted. The fluid used is Karo$\textsuperscript{TM}$ brand light corn syrup, with dynamic viscosity ($\mu$) of 1.229 kg m$^{-1}$s$^{-1}$, and density ($\rho$) of 1340 kg m$^{-3}$ at an ambient room temperature of 20$^{\circ}$C. 

The models are inserted flush on the top and bottom walls of an optically accessible plexiglass flow tank with dimensions as shown in Figure~\ref{setup}. The fluid is driven by means of a screw pump that is powered by a DC motor. To achieve parabolic inflow, the fluid entering and exiting the tank is passed through a set of flow straighteners consisting of a fine mesh. Note that in the schematic in Figure~\ref{setup}, $N$ is equal to 4 (as seen in the top view). This is merely for the purpose of illustration, as other variations of this parameter are examined as well (see Table \ref{exp_param} for details). The average inflow velocities with which the pump moves the fluid upstream of the models at the entrance is approximately $V_{exp} = 0.0002\ m/s$. Using this value as the velocity scale for the fluid properties mentioned earlier, a $Re_d$ value of approximately $0.001$ is achieved in these experiments, using the diameter $d_{pin} = 0.001\ m$ of the individual cylinders as the length scale, or $Re \approx 0.1$ using channel height $H$ as the length scale. Since the pins used in the design of the models are opaque, the flow through the layer could not be resolved using PIV since this method requires optical access to acquire information.\\

%
%
%
%

\subsection{Experimental Diagnostics}

In order to visualize the flow field and quantitatively determine the velocity, two-dimensional particle image velocimetry (PIV) measurements are conducted on the flow over the scale models in the flow tank. Particle image velocimetry is a non-intrusive, two or three dimensional technique that can be used to obtain instantaneous information on a flow field by recording and processing the single or multiple exposed images of tracer particles suspended in the fluid. The particle images are then processed using correlation based techniques to construct the velocity vector field of the fluid flow. Detailed reviews of PIV can be found in Adrian\cite{adrian:91arfm} and Willert \& Gharib\cite{willert:91expfluids}.

Measurements are made using time-averaged PIV. The laser sheet for the PIV measurements is generated from a 50 mJ double-pulsed Nd:YAG laser manufactured by Continuum Inc., which emits light at a wavelength of 532 nm with a maximum repetition rate of 15 Hz. The laser beam is converted into a planar sheet approximately 3 mm thick using a set of focusing optics. The laser sheet is located  in the $x$-$y$ plane along the center of the flow tank ($z$ = 21 mm in Figure~\ref{setup}). The time interval of separation between two images in an image pair is held constant at 0.1 s throughout all experiments. A 14 bit CCD camera (Imager Intense, LaVision Inc.) with a 1376$\times$1040 pixel array is used for capturing images. Uniform seeding is accomplished using 10 micron hollow glass spheres that are inserted in the flow tank and mixed to achieve a near homogenous distribution prior to each experiment. For processing the raw images, the software Davis 7.0 provided by LaVision Inc. is used. For each PIV run, 20 images are recorded for processing resulting in a minimum of 10 velocity vector fields from which to generate the mean flow field and statistics.

To check the appropriateness of this Brinkman model to the physical model, a least squares regression is used to find the values of $dp/dx$ and $\alpha$ that best approximate the experimentally determined velocity profile between the two layers as measured by the experiments described in Section~\ref{physical_model} . The equation describing the flow velocities between the two layers (\ref{eqn:Reg1gensol}) can be restated as follows:
\begin{equation}
u(y) = \frac{A}{2}y^{2} - Aby + C(\alpha),
\end{equation}

\noindent where $A = \frac{1}{\mu}\frac{dp}{dx}$, $b$ is half of the distance between the layers, and 
$C(\alpha)$ is a constant that depends upon $\alpha$.

\section{Results}
\label{sec3}

%
%
%
%

\subsection{Experimental Results}

\begin{figure}[H]
\centering{\includegraphics[width=6in]{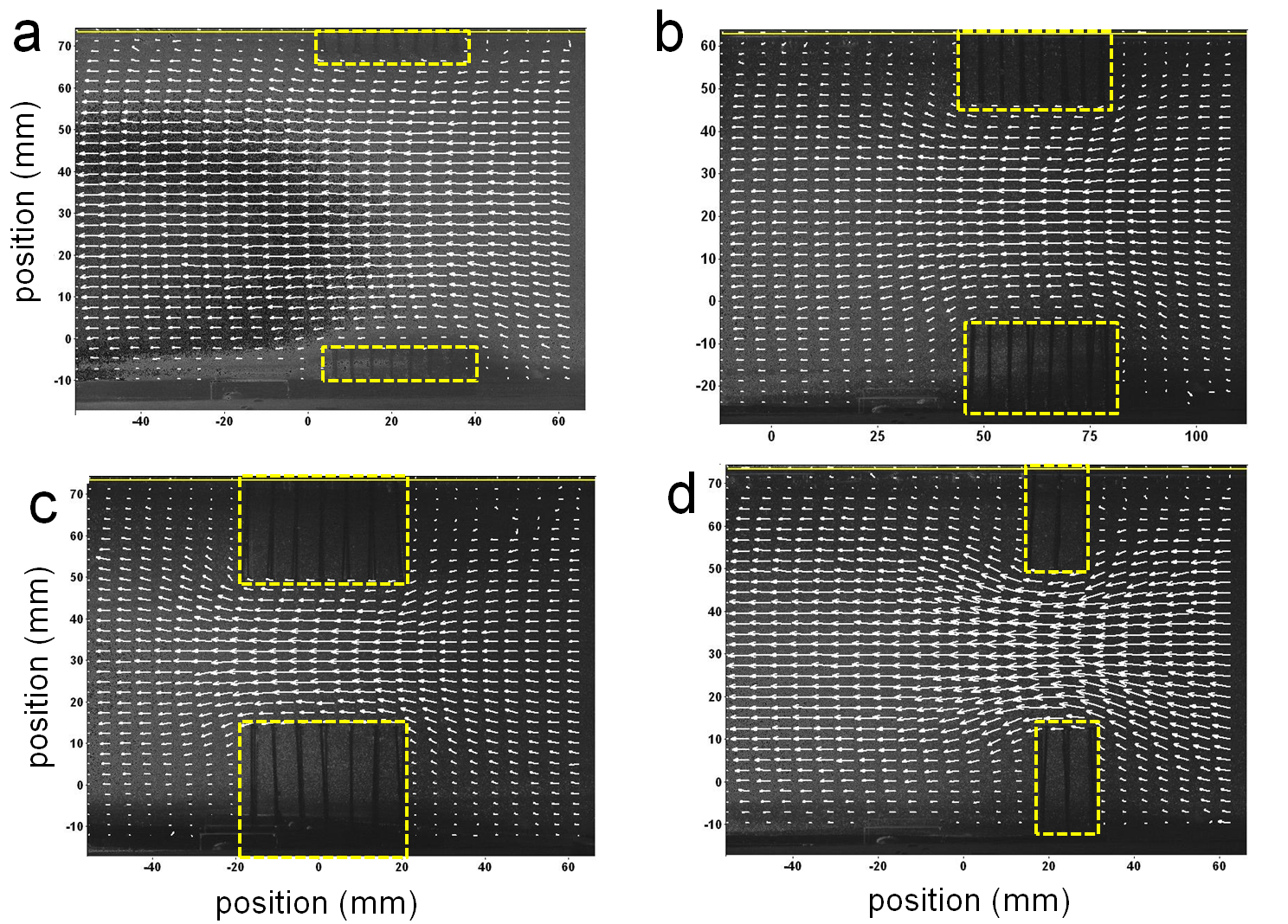}}
\caption{Some example velocity vector fields obtained from PIV measurements of the flow around the physical models from the side view. Flow visualization is also shown in Supplemental Movie 1. The lengths of the vectors are proportional to the magnitude of the velocity. Flow is from right to left. The background of the images are snapshots of the flow. The regions where the model pins are located are enclosed by yellow boxes.}
\label{PIV_glyco}
\end{figure}

Figure \ref{PIV_glyco} shows some example velocity vector fields obtained from PIV measurements of the flow around the physical models from the side view. For all cases, the diameter and spacing of the pins is kept constant at $d_{pin} = 0.001\ m$ and spacing$ = 0.005\ m$. For (a,b,c), $L = 0.043\ m$. The length of the layer was decreased in (d) to about $0.010\ m$. The height of the layers was varied from (a) $0.008\ m$, (b) $0.022\ m$, and (c,d) $0.028\ m$. The lengths of the vectors are proportional to the magnitude of the velocity. Flow is from right to left. The backgrounds of the images are snapshots of the fluid and pins. The regions where the model pins are located are enclosed by dashed boxes. In all cases,  the flow velocities increase within the space between the layers as seen by the longer vectors present in these regions. In general the velocity increases as the space between the layers decreases. The flow along the floor moves upwards as it approaches the clump of pins and then downwards as it moves past the pins. The flow between the arrays could not be resolved on account of the opacity of the pins. 

%
%
%
%

\subsubsection{Comparison of Experimental Results to 1D Theory}

\begin{figure}[H]
\centering{\includegraphics[width=6in]{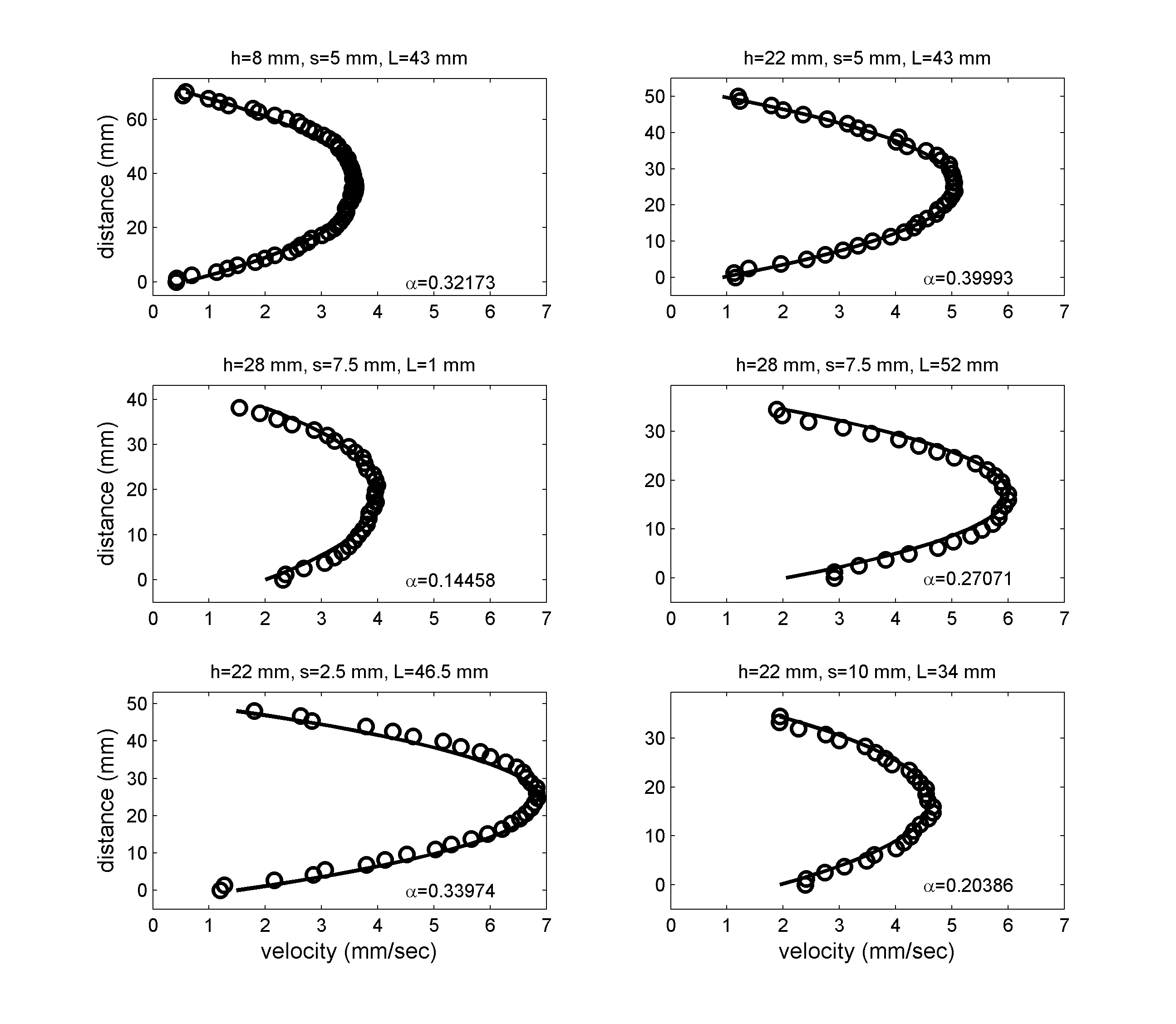}}
\caption{Experimentally and theoretically determined velocity profiles between two model layers. The experimental data is labeled with open circles, and the theoretical predictions are denoted with a solid line. The values of $\alpha$ have units of cm$^{-1}$. In the top row, the spacing between pins and layer length is fixed at 5 mm and 43 mm, respectively. The height is varied from 8 mm (left) to 22 mm (right). The middle rows shows the velocity profiles are shown between two layers with a constant height of 28 mm and spacing of 7.5 mm. The length of the layer was varied from 1 mm (left) to 52 mm (right). The bottom row shows profiles for a constant layer height of 22 mm and lengths of 46.5 and 34 mm. The spacing was varied from 2.5 mm (left) to 10 mm (right).}
\label{exp_theor}
\end{figure}

Figure~\ref{exp_theor} shows the experimentally measured velocity profiles between two layers compared to the profiles determined by the Brinkman model. As explained in the methods section above, the values of $dp/dx$ and $\alpha$ are determined using a least squares regression. The top panel shows velocity profiles for layer heights of 8 mm (left) and 22 mm (right). In both cases, the spacing between the pins is set to 5 mm and the length of the layer is 43 mm. Since the layer density is the same in both cases, one would predict that the value of $\alpha$ would remain fairly constant. $\alpha$ was determined to be 0.32173 mm$^{-1}$ for the shorter layer and 0.39993 mm$^{-1}$ for the taller layer. In the middle panel, the length of the layer is changed from 1 mm (left) to 52 mm (right), while the spacing and height is held constant (layer height$=28$ mm, spacing$= 7.5 mm$). Although the volume fraction of the layer is unchanged, the value of $\alpha$ drops nearly 50\% as the layer length is reduced to one row of pins. Since the theoretical model assumes fully developed flow between the layers, it is not surprising the model with $\alpha$ held constant would not accurately predict the flow profile. The bottom panel shows the flow profiles for variable layer densities, with layer spacings of $\{2.5, 10\}$ mm. As expected, the values of $\alpha$ for the models with a higher volume fraction ($\alpha = 0.33974$ mm$^{-1}$) is greater than that for the sparser models ($\alpha = 0.20386$ mm$^{-1}$).

To compare the values of hydraulic resistances between the model, experiment, and estimated values for capillaries, $\alpha$ will be expressed in nondimensional form. Let $U$, $L$, and $T = L/U$ be the characteristic length (height of chamber), velocity (peak velocity), and time scales, respectively. Define $u' = u^*/U, y' = y/L, x' = x/L, p' = p/(\rho U^2)$, and $\alpha' = \alpha L$. Then $x$-momentum equation (\ref{eqn:Brink1}) within the layer becomes:

\begin{equation}
\frac{d p^{\prime}}{d{x^{\prime}}} =  \frac{1}{Re} \frac{d^{2} u'}{d y'^{2}} -  \frac{(\alpha')^{2}}{Re} u',
\end{equation}


\noindent where ${\alpha^{\prime}}^{2}$ is the nondimensional, scaled resistance coefficient. To obtain the nondimensional resistance coefficient in Section 2 of the Results, the channel height (0.0006 cm) is multiplied by the values of $\alpha$ in cm$^{-1}$. This gives a range of $\alpha'^{2}$ from 1.44 to 562,500. Rapid changes in the flow profile are observed for $\alpha'^{2}$ on the order of 1000. To nondimensionalize the values of $\alpha$ that are experimentally derived in Section 3 of the Results, the channel height (8 mm) is multiplied by the values of $\alpha$ in mm$^{-1}$. This gives a range of $\alpha'^{2}$ from about 130 to 1025. Secomb et al. \cite{Seco:98} estimated that the dimensional hydraulic resistivity of rat mesentery was on the order of 10$^{8}$ dynes s/cm$^{4}$. Converting this value to the nondimensional resistance coefficient gives $\alpha'^{2} \approx 3600$. 

These results show that the flow patterns outside of the porous layer are well described by the Brinkman equation even when the layer is sparse or made up of clumps of relatively short lengths. Comparison of the experimental flow fields to the results reported by Leiderman \textit{et al.}\cite{Leiderman:2008} where the endothelial surface layer is modeled as a two-dimensional Brinkman `clump' also show good agreement.

%
%
%
%

\subsection{3D simulations of flow through arrays of cylinders}

%
%
%
%

\subsubsection{Effect of Re}

We varied the Reynolds number with $Re\in \{0.2, 1.0, 10\}$. This corresponds to diameter based Reynolds numbers $Re_d\in \{0.00625, 0.03125, 0.3125\}$. Over this range, our simulations suggest no significant change in average flow velocity in either direction perpendicular to the cylinder. However, average vertical flow velocity increases with $Re$ within the cylinder layer. This velocity increase is at its maximum in a plane located at approximately $28.8\%$ of the cylinder's height with velocity separation due to $Re$ disappearing at the base and the top of the cylinder. The difference in average vertical velocity scales linearly with the magnitude of the Reynolds number according to the equation $0.023612\log_{10}(Re_1/Re_2) = \Delta\mbox{avg}(u_z)$, where $\mbox{avg}(u_z)$ is the average magnitude of flow velocity in the $z$-direction within a plane located at $28.8\%$ of the cylinder's height.

\begin{figure}[H]
    \centering
    \includegraphics[width=0.6\textwidth]{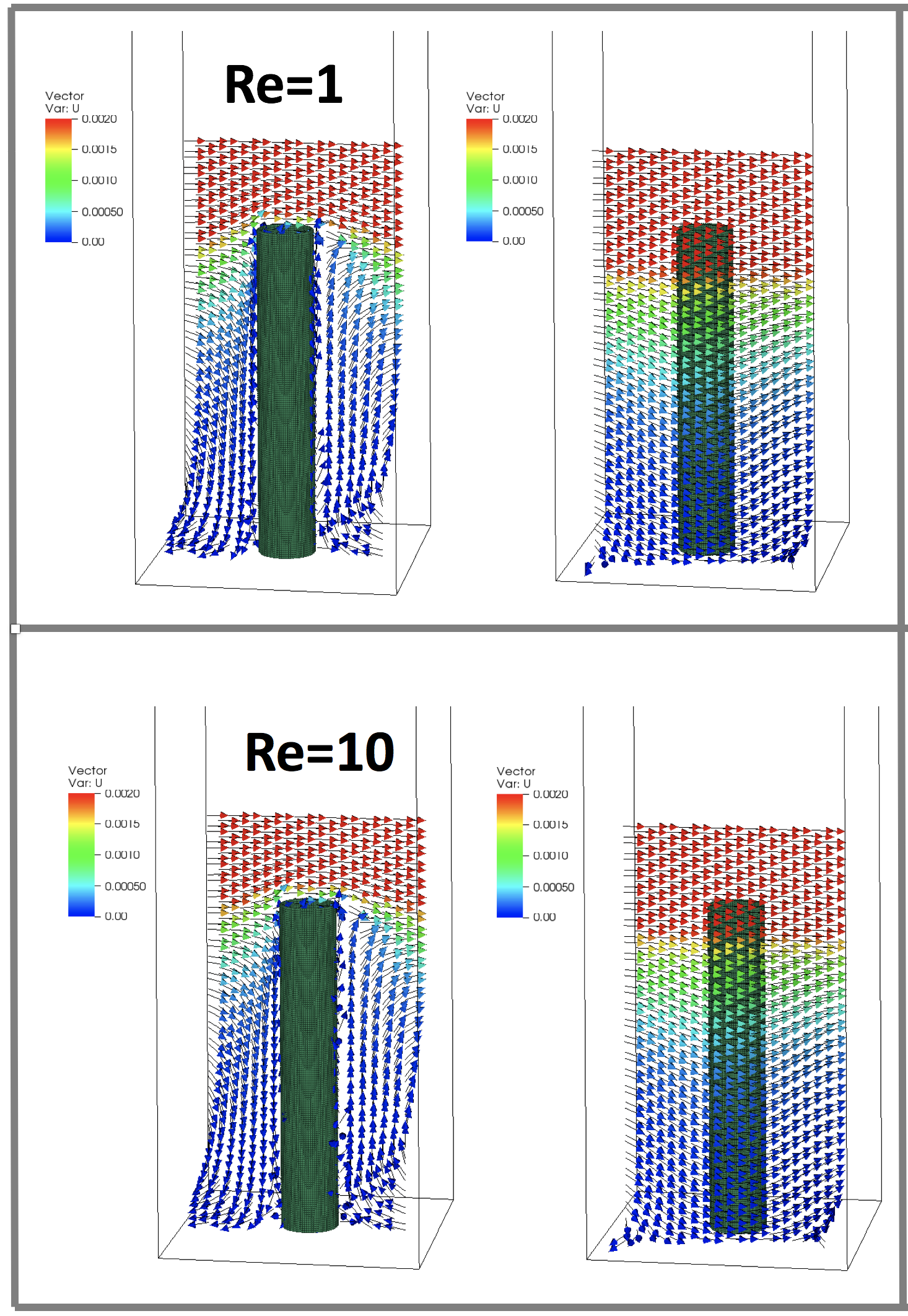}
    \caption{Velocity vectors along planes taken parallel to the direction of flow and through the center of the cylinder (left images) and between periodic cylinders (right images) at $Re = 1, 10$ ($Re_d = 0.03125,0.3125$). Reynolds numbers were changed by varying the dynamic viscosity. The height of the cylinder is set to 0.156, such that H/D = 5 and G/D = 4.}
    \label{fig:Re_flow}
\end{figure}

Figure \ref{fig:Re_flow} shows a sample flow through the cylinders at $Re\in\{1,10\}$. The arrows show the direction of flow and the color corresponds to the magnitude. Velocity vectors along planes are taken parallel to the direction of flow and through the center of the cylinder (left images) and between periodic cylinders (right images). Note that the height is set to 0.156 (such that H/D = 5 and G/D = 4). The flow between the cylinders is typically quite small relative to the free stream velocity.

%
%
%
%

\subsection{Effect of the number of cylinders}

\begin{figure}[H]
    \centering
    \includegraphics[width=0.9\textwidth]{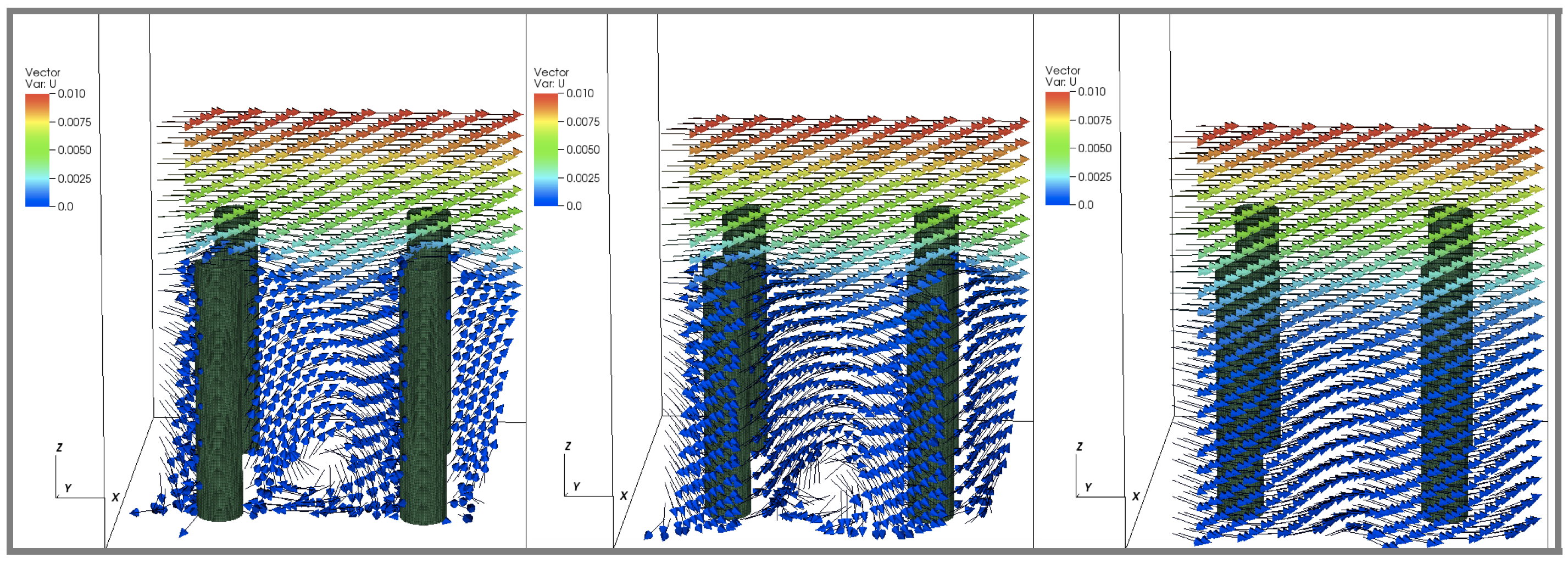}
    \caption{Velocity vectors taken within planes parallel to the direction of flow at $Re = 1$ ($Re_d=0.03125$). The planes are taken through the center of the cylinder (left), one diameter from the center of a cylinder (middle), and half way between rows of cylinders (right). The height of the cylinders were set to 0.156, such that H/D = 5 and G/D = 4.}
    \label{fig:No_flow}
\end{figure}

To consider the effect of the periodic boundary conditions, we varied the number of cylinders within the periodic domain as 1, 4, and 16 cylinders. The effective spacing between the cylinders was kept constant such that G/D=4, and the size of the domain was set to 0.125, 0.25, and 0.5. The height of the cylinders was kept fixed as $h=10/64$ such that H/D = 5. The $Re$ was set to 0.2, 1, and 10 for each number of cylinders (such that $Re_d = 0.03125,0.3125$). Across these simulations, the number of cylinders did not appear to have any significant effect on the average flow velocity profile. Some differences were observed in the small scale local velocity fields within the layer. Figure \ref{fig:No_flow} shows velocity vectors within planes taken parallel to the direction of flow at $Re = 1$. The planes are taken through the center of the cylinder (left), one diameter from the center of a cylinder (middle), and half way between rows of cylinders (right). Slow moving regions of rotation are observed near the base of the cylinders. If only average flow profiles are desired, periodic boundary conditions with a single cylinder are sufficient to model an array of larger size within the specified parameter space. If details of the flow near the surface are needed, the results suggest that larger periodic domains are required.

%
%
%
%

\subsubsection{Effect of height}

\begin{figure}[H]
    \centering
    \includegraphics[width=0.9\textwidth]{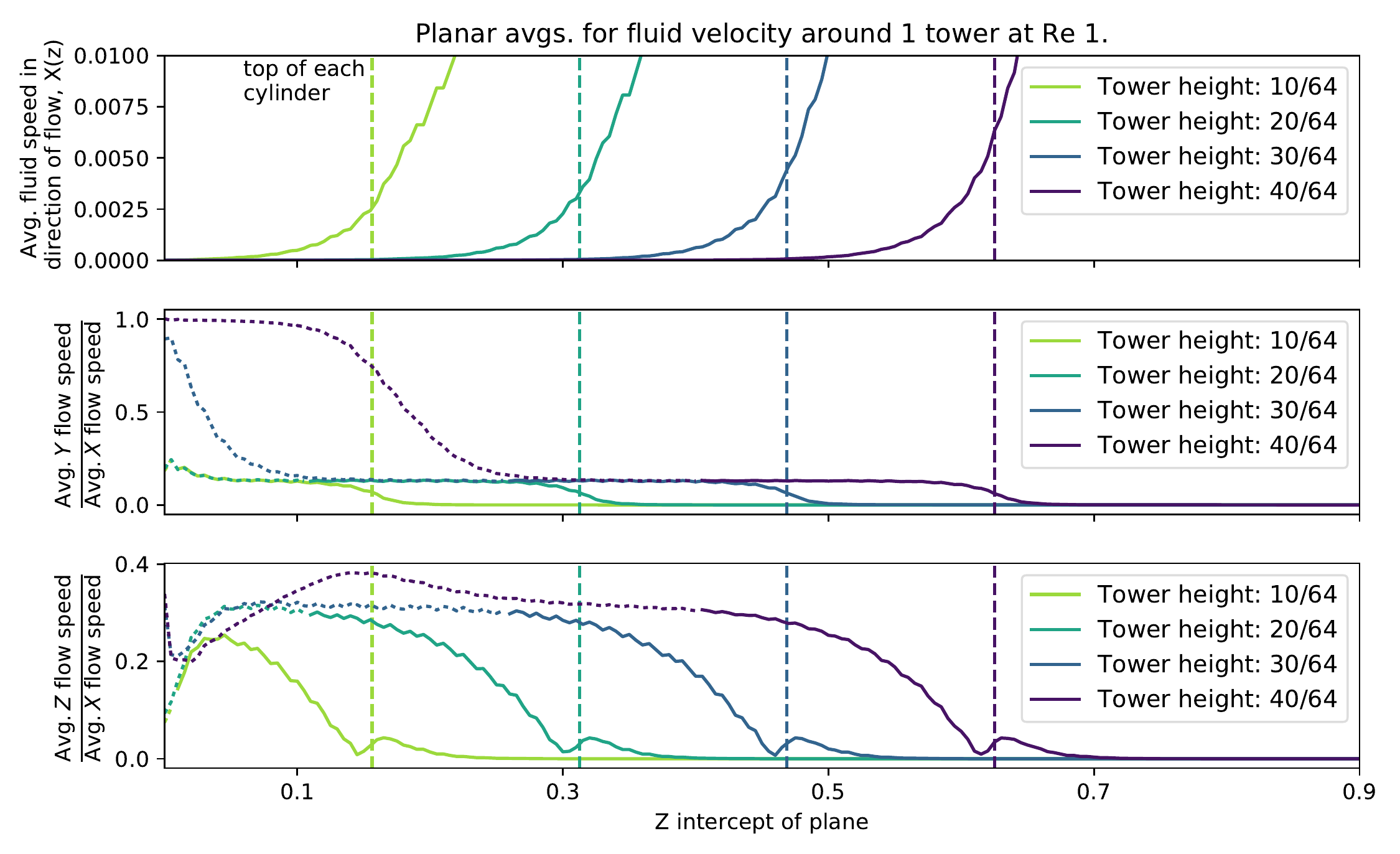}
    \caption{Magnitude of flow velocity averaged across planes perpendicular to the $z$ axis for four different cylinder heights. The top of each cylinder is plotted as a vertical dashed line. A dotted line has been used in the $y$ and $z$ plots where the average fluid velocity in the direction of the flow ($x$) falls below $1e-5$.}
    \label{fig:height_planes}
\end{figure}

To consider the effect of cylinder height, we set the height equal to 10/64, 20/64, 30/64, and 40/64 (corresponding to $H/D = 5, 10, 15, 20$). The domain size was fixed at 0.125 and 0.25 with a single cylinder and four cylinders, respectively. The $Re$ was set to 0.2, 1, and 10. Figure \ref{fig:height_planes} shows the magnitude of the velocity average across the $xy$-plane as a function of height within the cylinder layer. The average fluid speed profile appears consistent between all cylinder heights, although for taller cylinders there is less distance between the top of the cylinder and the maximal average flow rates, i.e., there is greater shearing at the top of the taller cylinders. Furthermore, near the bottom of the domain there is more fluid motion in the $y-direction$, due to the non-zero and zero shearing in the $x-direction$ at the top and bottom of the domain, respectively. As you measure the flow components moving towards the top of the domain, the fluid begins to move for uniformly in the $x-direction$, with little motion in the $y$. However, interestingly, the $y:x$ component ratio of the average flow speeds, shows that as the $x$-component begins to dominate, it temporarily tapers down to a constant value, $\sim 0.2$, until you measure high enough in the domain, in which case, the $x$-motion dominates and there appears little or no motion in the $y-$direction. Moreover, for the ratio of $z:x$ components of the average velocity, a similar trend occurs, where the $x-$average velocity component dominates. However, rather than the ratio monotonically decrease to zero, as in the $y:x$ case, for each tower height, the ratio hits a relative minimum, shortly increases, and then tapers off to zero as you measure further up the height of the domain. In each case, the relative maximum of the ratio of $z:x-$average velocities occurs for heights slightly about the height of each cylinder. 

%
%
%
%

\subsubsection{Effect of spacing}

To consider the effect of cylinder spacing, we fixed the height at 10/64 and set the domain width to 0.125, 0.25, 0.5, and 1. This provided a range of gap to diameters such that $G/D = 4, 8, 16, 32$. $H/D$ was kept fixed at 5. The $Re$ was set to 0.1, 1, and 10 ($Re_d \in \{0.003125, 0.03125, 0.3125$\}). 

\begin{figure}[H]
    \centering
    \includegraphics[width=0.9\textwidth]{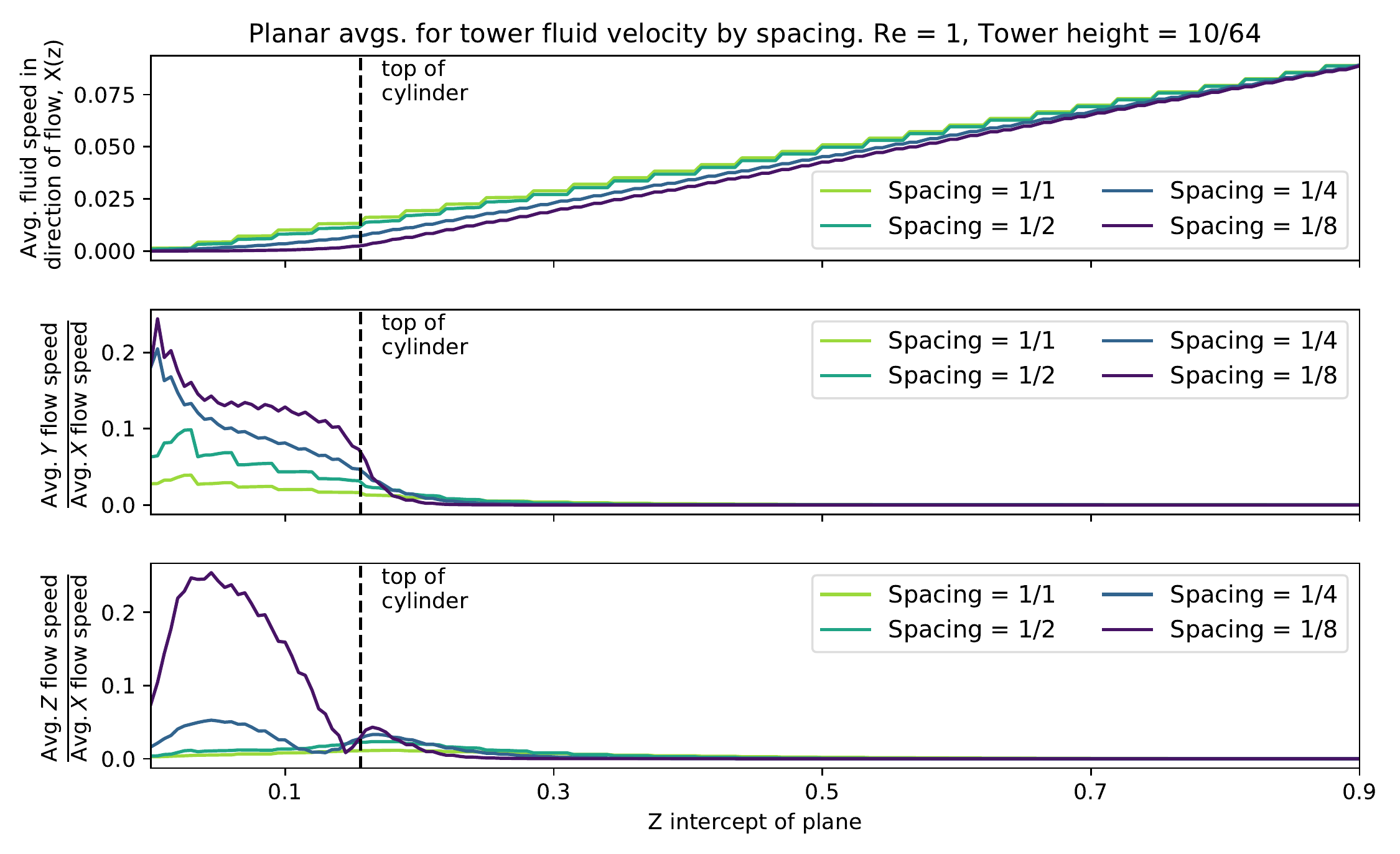}
    \caption{Magnitude of flow velocity averaged across planes perpendicular to the $z$ axis for different cylinder spacing. The top of the cylinders is plotted as a vertical dashed line.}
    \label{fig:spacing_planes}
\end{figure}

Figure \ref{fig:spacing_planes} shows the magnitude of the velocity average across the $xy$-plane as a function of effective spacing. It was found that as you measure the average flow speed in the $x-$direction as a function of $z$, it increases in almost a linear fashion above the top of the cylinder. Furthermore, high enough above the tower, the ratios of the $z:x$ and $y:x$ components of average velocity taper to zero, while for $z-$cross-sections below the top of the tower, there are non-linear relationships between those ratios. Moreover, when the spacing is reduced, there is more average fluid motion in the $y-$ and $z-$directions within the height of the tower. In the case of the $z:x$ ratio, there exist clear local maximum for spacings of $1/8$ and $1/4$, one within the height of the cylinder and another slightly above the top of the cylinder. This is in contrary to the $y:x$ case, where the ratio more monotonically tapers towards zero. 

%
%
%
%

\subsection{Brinkman vs Explicit Treatment of Cylinders}

We compare the results of the averaged flow as a function of height from the three-dimensional simulations to the 1D analytical model. To begin, we average the flow across the $xy$-plane at each height. We then fit the numerical data to the Brinkman model using nonlinear least squares to find the best choice of the porosity.

\begin{figure}[H]
    \centering
    \includegraphics[width=0.8\textwidth]{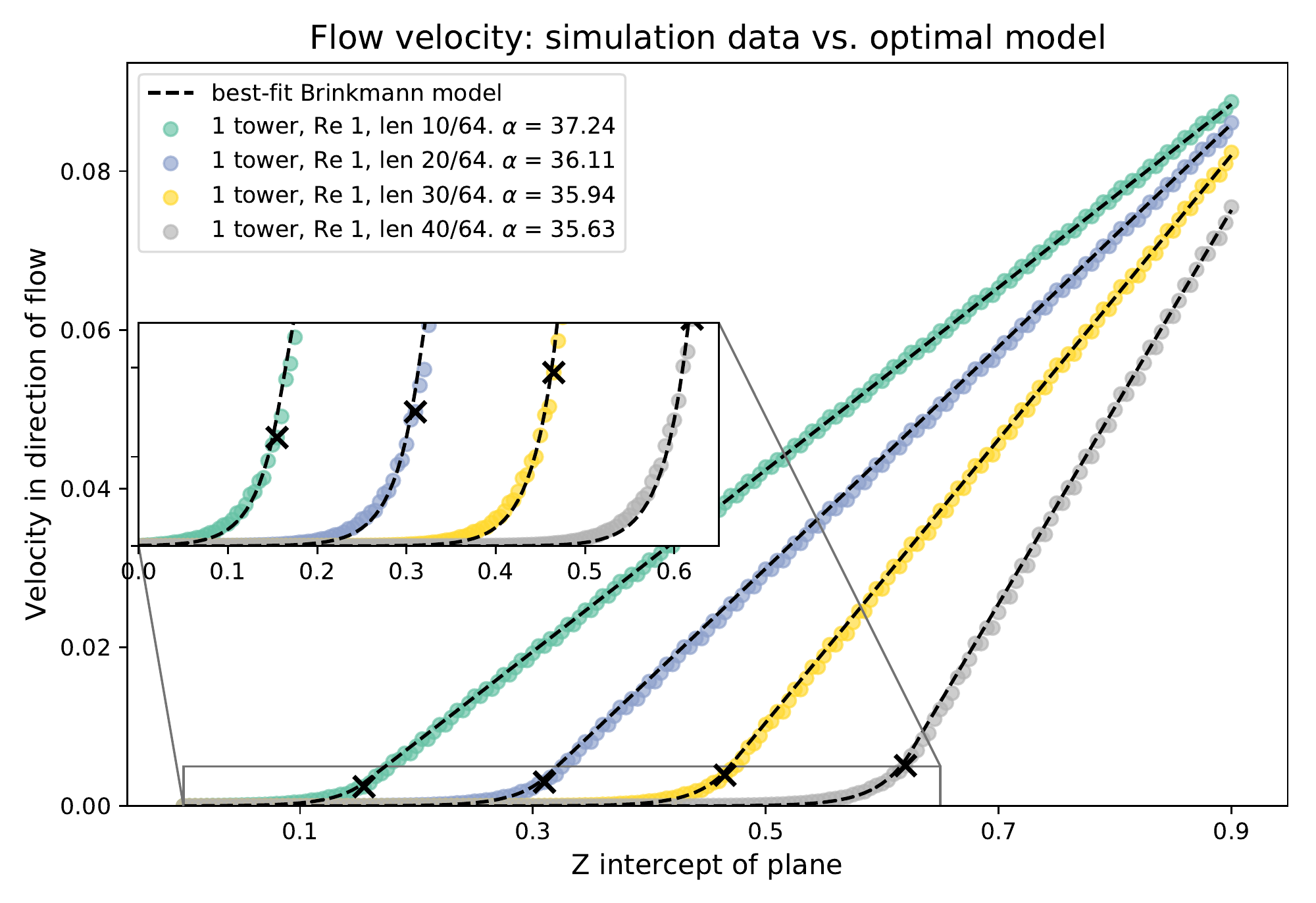}
    \caption{Average velocity in the direction of the flow for various cylinder heights, compared with the best fitting Brinkman model. The top of each cylinder is marked with a black X. The best choice of porosity is given for each cylinder height in the figure legend.}
    \label{fig:brinkman}
\end{figure}

Figure \ref{fig:brinkman} shows a sample of the averaged velocity magnitude from the numerical data vs. height compared to the analytical model. The best fit porosity coefficients are included. The $3D$ simulations and analytical results are consistent and give the same flow profiles for the $x-$component of the average velocity and height, $z$. Moreover, from  finding the best fit porosity coefficient for the analytical model, it appears that as the height of the tower increases, the porosity coefficient decreases. 

%
%
%
%

\section{Discussion}
\label{sec4}

The flow through biological filtering layers can be highly three-dimensional. In many porous models, these dynamics are not properly captured. While bulk flow may appear to be in only the $x-$direction, transient flows in the $y-$ and $z-$directions are important as they can enhance nutrient and chemical exchange and alter settlement. Our $3D$ simulation results indicate that the amount of flow through such layers depends nonlinearly on both $Re$, layer height, and spacing between successive cylinders. The Brinkman model may only be sufficient for describing flow above the layer and giving bulk flow average estimates flow within the porous layer. To fully resolve the fluid motion in these other directions fully coupled FSI is necessary, as illustrated by our results. 

Moreover our bulk flow profiles obtained from the physical and numerical models do agree with previous theoretical work. With regard to the endothelial surface layer, the flow rates through the layer are small, and the majority of shear stress is felt on the luminal side \cite{Seco:01a, Dami:98, Feng:00}. In regards to trichomes, our results suggest that flow near the leaf surface is slow, and velocities in the vertical direction may be on the order of flow in the tangential direction. Trichomes may also operate near the transition between acting as a leaky and a solid layer. For tiny insects, our results support previous work that the wing's bristled morphology operates near the leaky-to-solid transition. Finally, the results of this paper show that the Brinkman model is a reasonable approximation of the average flow within and above micro- and mesoscale porous layers even if the volume fraction of obstacles (e.g. core proteins, trichomes, bristles) is relatively low. Assuming that the layer is relatively homogeneous and of reasonable length relative to the height of the layer, the one-dimensional Brinkman model can be used to obtain analytical solutions of the averaged flow profiles and shear stresses. 

In this initial study, we have resolved the three-dimensional flows through arrays of rigid, evenly spaced cylinders at $Re<10$. In the biological world, hairs, bristles, and other structures are more complex. Future work should explore the role that flexibility plays on flow. Furthermore, variations in shape, such as those seen in trichomes, may alter the local flow significantly. Accordingly, additional studies are needed to quantify the effect of shape on flow. Finally, we limit our study to $Re<10$, where strong inertial effects such as vortex separation are not observed. Although previous studies have considered flow through obstacles at large scales ($Re>1000$ \cite{Kim:Thesis,Bazilevs:2014,Kinzel:2015}), additional work is needed to map up transitional flows for $10<Re<1000$.

%
%
%
%

\section{Acknowledgements}
The authors would like to thank the following grant sources:  Army Research Office Staff Research Grant to Pasour; NSF DMS Math Biology and CBET Fluid Dynamics CAREER Award 1151478, NSF DMS Math Biology and CBET Fluid Dynamics 1022802 to Miller; NSF CBET Fluid Dynamics 1512071 grant to Santhanakrishnan.

\bibliographystyle{elsarticle-num}

\bibliography{sample}

\end{document}